\documentclass{aastex61}
\usepackage{graphicx,epstopdf,epsfig}
\usepackage{natbib}
\usepackage{csquotes}

\shortauthors{Mazumder et al.}

\begin{document}

\title{Solar Cycle Evolution of Filaments over a Century: Investigations with the Meudon and McIntosh Hand-drawn Archives}

\author{Rakesh Mazumder}
\correspondingauthor{Dipankar Banerjee}
\email{dipu@aries.res.in}
\affiliation{Aryabhatta Research Institute of Observational Sciences, 
Manora peak, Nainital-263 001, Uttarakhand, India}
\affiliation{Center of Excellence in Space Sciences India, Indian Institute of Science Education and Research Kolkata, Mohanpur 741246, West Bengal, India}

\author{Subhamoy Chatterjee}
\affiliation{Southwest Research Institute, 1050 Walnut St \#300, Boulder, CO 80302, USA}

\author{Dibyendu Nandy}
\affiliation{Center of Excellence in Space Sciences India, Indian Institute of Science Education and Research Kolkata, Mohanpur 741246, West Bengal, India}
\affiliation{Department of Physical Sciences, Indian Institute of Science Education and Research Kolkata, Mohanpur 741246, West Bengal, India}

\author{Dipankar Banerjee}
\affiliation{Aryabhatta Research Institute of Observational Sciences, 
Manora peak, Nainital-263 001, Uttarakhand, India}
\affiliation{Indian Institute of Astrophysics, Koramangala, Bangalore 560034, India}
\affiliation{Center of Excellence in Space Sciences India, Indian Institute of Science Education and Research Kolkata, Mohanpur 741246, West Bengal, India}

\begin{abstract}
Hand-drawn synoptic maps from the Meudon Observatory (1919 onwards) and the McIntosh archive (1967 onwards) are two important sources of long-term, manually recorded filament observations. In this study, we calibrate the Meudon maps and subsequently identify filaments through an automated method. We extract physical parameters from this filament database and perform a comparative study of their long-term evolution focusing on the cotemporal period of McIntosh and Meudon observations. The spatio-temporal evolution of filaments manifests in the form of a filament butterfly diagram, indicating further that they are intimately related to the large-scale solar cycle. Physical descriptors such as the number and length of filaments, which are tracers of solar surface magnetic field, have cycles which are phase-locked with the $\sim$ 11 year sunspot cycle. The tilt angle distribution of filaments -- both near or distant from active region locations -- indicates that their origin is due to either large-scale surface magnetic field or inter-active region field evolution.
This study paves the way for constructing a composite series of hand-drawn filament data with minimal gaps stretching the time span of solar filament observations to a century. On the one hand, this would serve as useful constraints for models of magnetic field emergence and evolution on the Sun's surface over multiple solar cycles, and on the other hand, this filament database may be used to guide the reconstruction of filament-prominence associated eruptive events before the space age.    

\end{abstract}


\section{Introduction}

Filaments are dark thin elongated structures on the solar disk. The same structure appears bright when seen above the limb and is called prominence for historical reasons. Filaments are dense cool plasma structures suspended in the solar corona. Filament eruptions often produce coronal mass ejection (CMEs) which create hazardous space weather \citep{2000ApJ...537..503G,2003ApJ...586..562G,2004ApJ...614.1054J,Sinha2019}. So theoretical and observational study of filaments is important for space weather. Though filaments form in the lower solar atmosphere they are related to magnetic flux emergence and evolution on the solar surface and hence reveal valuable information about the large-scale evolution of solar magnetic fields and the sunspot cycle. Filaments always form over polarity inversion lines (PILs) \citep{1998SoPh..182..107M,1972RvGSP..10..837M,1983SoPh...85..215M}, the line separating two opposite signed magnetic patches on the solar surface. \cite{2014SoPh..289..821K} studied the criteria for the formation of the filaments over a PIL. Depending on the position of the filaments in the solar disk they are divided into different categories \citep{1923CompRend...176..950,1928AnnObsParis...VI..1,1948AnnObsParis...6..7} namely active region filaments, quiescent region filaments, and polar filaments. \cite{1994ASIC..433..303M} categorized filaments as dextral and sinistral by their chiral properties. They reported the northern hemisphere was dominated by dextral filaments whereas the southern hemisphere was dominated by sinistral filaments \citep{2003ApJ...595..500P,2005SoPh..228...97B,1996ApJ...460..530P}.

Different features on the Sun are manifestations of solar magnetic field distribution and are crucial in gauging solar magnetic field evolution in the past before direct measurement of the large-scale solar magnetic fields became possible after the invention of the magnetograph. Solar filaments are special as they appear across latitudes depending on the phase of solar cycles and studies indicate the importance of their long-term evolution for illuminating the polar field build-up process \citep{{2017ApJ...849...44C},{doi:10.1029/2019EA000666},{2018ApJ...868...52M},{2021ApJ...909...86X}}. There are several studies on automated, semi-automated detection of filaments from solar images \citep{{2005SoPh..227...61F},{2015ApJonka..221...33H},{2005SoPh..228...97B},{2011SoPh..272..101Y},{2013SoPh..286..385H},{2014SoPh..289.2503S},{2007ASPC..368..505A}} and synoptic maps \citep{SWE:SWE20537,2017ApJ...849...44C}. However, historically significant effort has been made to detect and produce catalogues of solar features such as sunspots, plages, filaments with manual intervention of experienced human operators \citep{2020arXiv200102939B}. Careful evaluation of these enables us to extract useful information and make a composite of different manual catalogues through comparison of results at the overlapping period.

McIntosh archive is one such hand-drawn archive that contains spatial information of sunspots, plages, filaments polarity inversion lines (PILs), and coronal holes. \cite{2018ApJ...868...52M} performed a detailed analysis of filaments' properties from McIntosh hand-drawn archive and found that the filaments are uniformly distributed in longitude in the Sun. However, they are not uniform in latitude. Latitude distribution of filaments shows a bimodal nature with two peaks between $10^{\circ}$ to $30^{\circ}$ \citep{2018ApJ...868...52M,2015ApJS..221...33H}. The temporal variation of latitude distribution of filaments renders a butterfly pattern \citep{1948AnnObsParis...6..7,1994A&A...290..279M,2015ApJS..221...33H,2016SoPh..291.1115T,2018ApJ...868...52M} with a much wider distribution in latitude (-$80^{\circ}$ to +$80^{\circ}$ degrees) in comparison with the butterfly diagram of sunspots. \citet{2010MNRAS.405.1040L} reported the drift velocity of filaments in latitude is quite different from that of sunspots' latitudinal drift velocity. The drift velocity also depends on the amplitude of the activity cycle \citep{2001SoPh..202...11M}.
Another distinguing feature of fialments' butterfly diagram is the migration to the pole during the maximum of the solar cycle \citep{2018ApJ...868...52M,2015ApJS..221...33H, 2021ApJ...909...86X}, which is famously known as ``polar rush". \cite{2018ApJ...868...52M} reported the coincidence of the timing of this filaments' migration to the pole with polar field reversal \citep{1983SoPh...85..215M,1972RvGSP..10..837M}. Measurement of polar field is very challenging due to the line of sight effect in the measurement of magnetic field in the limb. Hence the timing of filaments migration to the pole can be used as a proxy to the timing of polar field reversal.

Tilt angle of active regions play a crucial role in the solar cycle evolution of magnetic fields. Specifically, the tilt angle distribution of active regions determine the efficiency of conversion of toroidal magnetic field to poloidal magnetic field in the solar dynamo mechanism \citep{1961ApJ...133..572B,Bhowmik2018}. Since filaments always form over polarity inversion lines (PILs) \citep{1998SoPh..182..107M,1972RvGSP..10..837M,1983SoPh...85..215M} the tilt angle of the filaments are perpendicular to the tilt angle of the active region. According to Joy's law \citep{1919ApJ....49..153H} the tilt angle of active region increase from the equator to pole. Thus the tilt angle of filaments being perpendicular to the active region decreases from equator to pole. The average tilt angle of filaments is reported to decrease from equator to pole by \cite{2016SoPh..291.1115T} and \cite{2018ApJ...868...52M}.  Variation of mean tilt angle of filaments with the solar cycle is also reported \citep{2016SoPh..291.1115T,2019RAA....19...80M}. Note also that the tilt angle of solar active region is negative in the northern hemisphere and positive in southern hemisphere according to Hale's polarity law \citep{1919ApJ....49..153H}. Active region filaments form in the polarity inversion line of the active region \citep{2008SoPh..248...51M}. The tilt angle of filaments is perpendicular to the active region tilt. So tilt angle of filaments is expected to be positive in northern hemisphere and negative in the southern hemisphere. However, \cite{2018ApJ...868...52M} reported filaments tilt to be predominantly negative in the northern hemisphere and positive in the south. \cite{2018ApJ...868...52M} and \cite{2019RAA....19...80M} speculated the main origin of the filaments is the evolution of large-scale magnetic field structures \citep{1983SoPh...85..215M,1989SoPh..123..367M,2000ARep...44..759M} as opposed to localized active regions -- a hypothesis that was motivated by this anomalous tilt angle distribution of filaments.

Generation of magnetic fields in the Sun is not symmetric in two hemispheres \citep{Bhowmik2019}. Naturally features like sunspots, filaments, plages, etc., which are magnetic in nature show north-south asymmetry. This north-south asymmetry of these features also varies with time.  \cite{2003NewA....8..655L,2010NewA...15..346L,2015RAA....15...77K,2015ApJS..221...33H} reported the asymmetry in filament number in different phases of solar cycle. \cite{2018ApJ...868...52M} reported the north-south asymmetry in filament lengths as well.

In this study, we perform a detailed analysis of solar filaments from Meudon hand-drawn archives to understand the origin and long-term behavior of filament. We compare these results with those from the McIntosh hand-drawn archive over their overlapping period. On the one hand, this enables us to corroborate hand-drawn observational records by different observers and on the other hand, this allows us to check for consistency in the long-term spatio-temporal evolution of filament properties. We describe the data-sets used for this study in section~\ref{data} and illustrate the calibration and detection methods in sections~\ref{calib_data} and \ref{fila_detect} respectively. We present and discuss the results in section~\ref{res_disc} and conclude in section~\ref{concl}.

\section{Hand-drawn Data Sources}\label{data}
\subsection{Meudon Synoptic Maps}
Meudon Observatory, France has archived a large set of hand-drawn synoptic maps spanning over the period 1919-2003 \citep{1948Paris..Obs..P,1994A&A...290..279M,1998ASPC..140..197M} \footnote{available at \url{ http://bass2000.obspm.fr/lastsynmap.php}}. They depict different magnetic structures such as sunspots, plages, and filaments. This dataset currently stands among one of the largest and oldest historical filament archives.  Though filament catalogue is available from the Observatory containing filament locations and labels, a detailed study of filament morphological parameters requires segmentation of the filaments. The upper left panel of Figure~\ref{carr_sample_hist} shows a representative Carrington map from Meudon database. The rightmost panel of Figure~\ref{carr_sample_hist} is the histogram of available observation days in Meudon data.

\subsection{McIntosh Synoptic Maps}
\begin{figure}
\centering
\includegraphics[scale=1.2,angle=90]{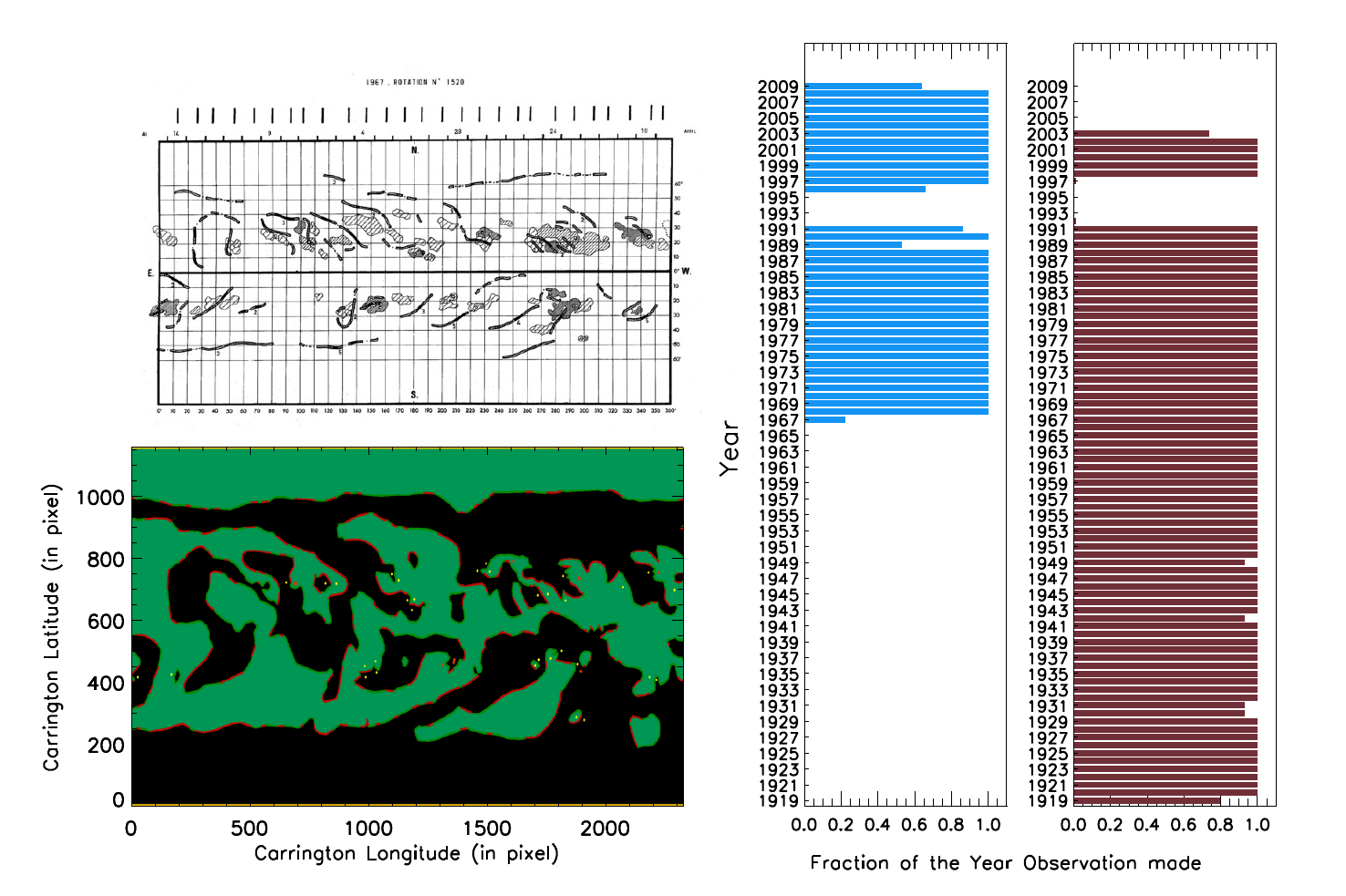}
\caption{Representative Carrington maps and map availability from Meudon and McIntosh database.  The upper left panel shows a Carrington map (rotation 1520) from Meudon database.  The lower left panel depicts a Carrington map for same rotation from McIntosh database. The middle panel shows yearly fraction of available observation days in McIntosh data as a function of time. The rightmost panel shows the yearly fraction for Meudon data.}
\label{carr_sample_hist}
\end{figure}

A scientist, Patrick McIntosh from NOAA's Space Environment Center at Boulder, created an archive of hand-drawn Carrington maps from 1967 April to 2009 July using both satellite- and ground-based observations, available at that time. These maps then scanned, digitized, and archived in NOAA/NCEI and made available both as images and `fits' format \footnote{\url{https://www2.hao.ucar.edu/mcIntosh-archive/four-cycles-solar-synoptic-maps}} by McA (McIntosh Archive) project (a Boston College/NOAA/NCAR collaboration, funded by the NSF). The Carrington maps in this archive contain filaments, polarity inversion lines (PILs), sunspots, plages, and coronal holes. The data cover solar cycle 20-23 (from 1967 April to 2009 July) with three data gaps one from 1974 June to 1978 July, another from 1991 October to 1994 January, and the last one from 1994 April to 1996 May. We have used level 3 `fits' files from this archive for our study.  The bottom left panel of Figure~\ref{carr_sample_hist} shows a representative Carrington map from McIntosh Archive. The middle panel of Figure~\ref{carr_sample_hist} is the histogram of available observation day in McIntosh data.

\section{Data Calibration}\label{calib_data}
\begin{figure}
\centering
\includegraphics[scale=0.9,angle=0]{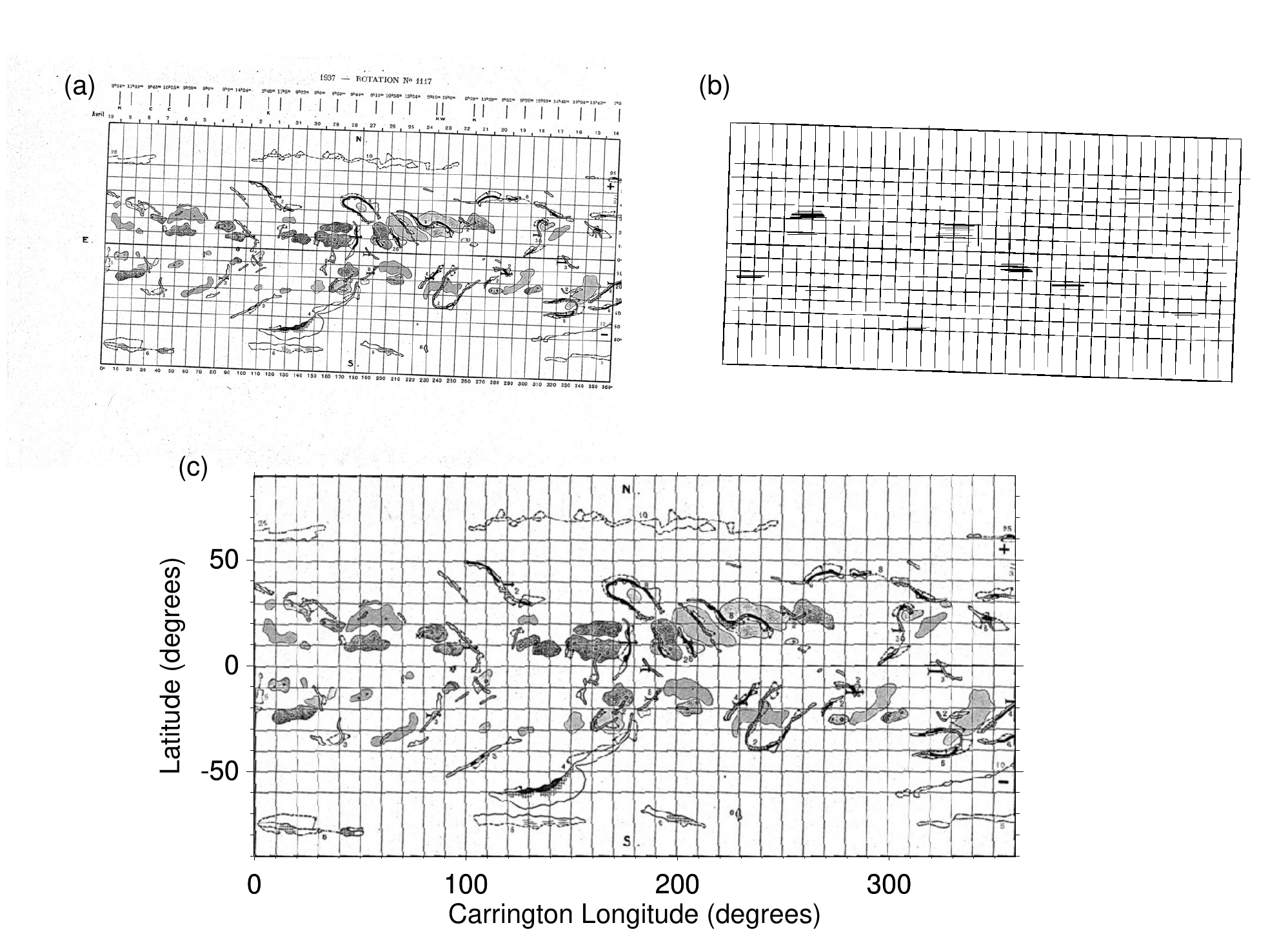}
\caption{Steps of calibration for Meudon hand-drawn Synoptic maps. (a) Map from rotation 1117 in original form; (b) Detected grids; (c) Tilt corrected, cropped and size normalised final map.}
\label{calib}
\end{figure}

Scanned hand-drawn Synoptic maps \citep{2016SoPh..291.1115T} from Meudon Observatory come in different sizes, origins and orientations. Thus, before filament detection, the maps need to be brought to the same origin, orientation and pixel scale. The calibration procedure is described in detail as below:
\begin{enumerate}
    \item We convert all the scanned Meudon maps (Figure~\ref{calib}a) in `jpeg' format to binary maps converting all the pixels lying below intensity 200 to 1’s and rest to 0’s.  The 1’s include the grids, filaments, and plages present in original maps.
    \item To separate the grids from the rest of the regions, we  morphologically erode the binary images \citep{sonka2014image} 3 times and subsequently dilate 3 times with horizontal and vertical structure functions having sizes as $\frac{1}{80}$ times the image x-dimension. The repeated erosion and dilation with horizontal structure-function separates the horizontal grids and the same with the vertical structure-function separates vertical grids from the rest of the images. A combined map of vertical and horizontal grids are shown in Figure~\ref{calib}b.
    \item From one of the horizontal grids we calculate the tilts of the maps and subsequently derotate the maps to make the equator horizontal.
    \item After rotation correction, we calculate the median x-coordinates from left-most, right-most vertical grids. We also calculate median y-coordinates from bottom-most and top-most horizontal grids. Through these x and y coordinates, we crop rectangular region of all the mapsand rebin them to the same size of 400 pixels (y) $\times$ 800 pixels (x) as depicted in Figure~\ref{calib}c.
    
\end{enumerate}
We have made all the gray scale and coloured calibrated images public.
\footnote{available at \url{https://github.com/rakeshmazumder/calibrated_filament_data}}
\pagebreak
\section{Filament Detection}\label{fila_detect}
Filaments in Meudon maps are recorded without color and with color for the rotations before 1823 and after 1823 respectively. We apply four different automated techniques to detect the filaments from four different ranges of Carrington Rotations (CRs) as described below.
\pagebreak
\subsection{Rotation 876-1823}
\begin{figure}
\centering
\includegraphics[scale=0.9,angle=0]{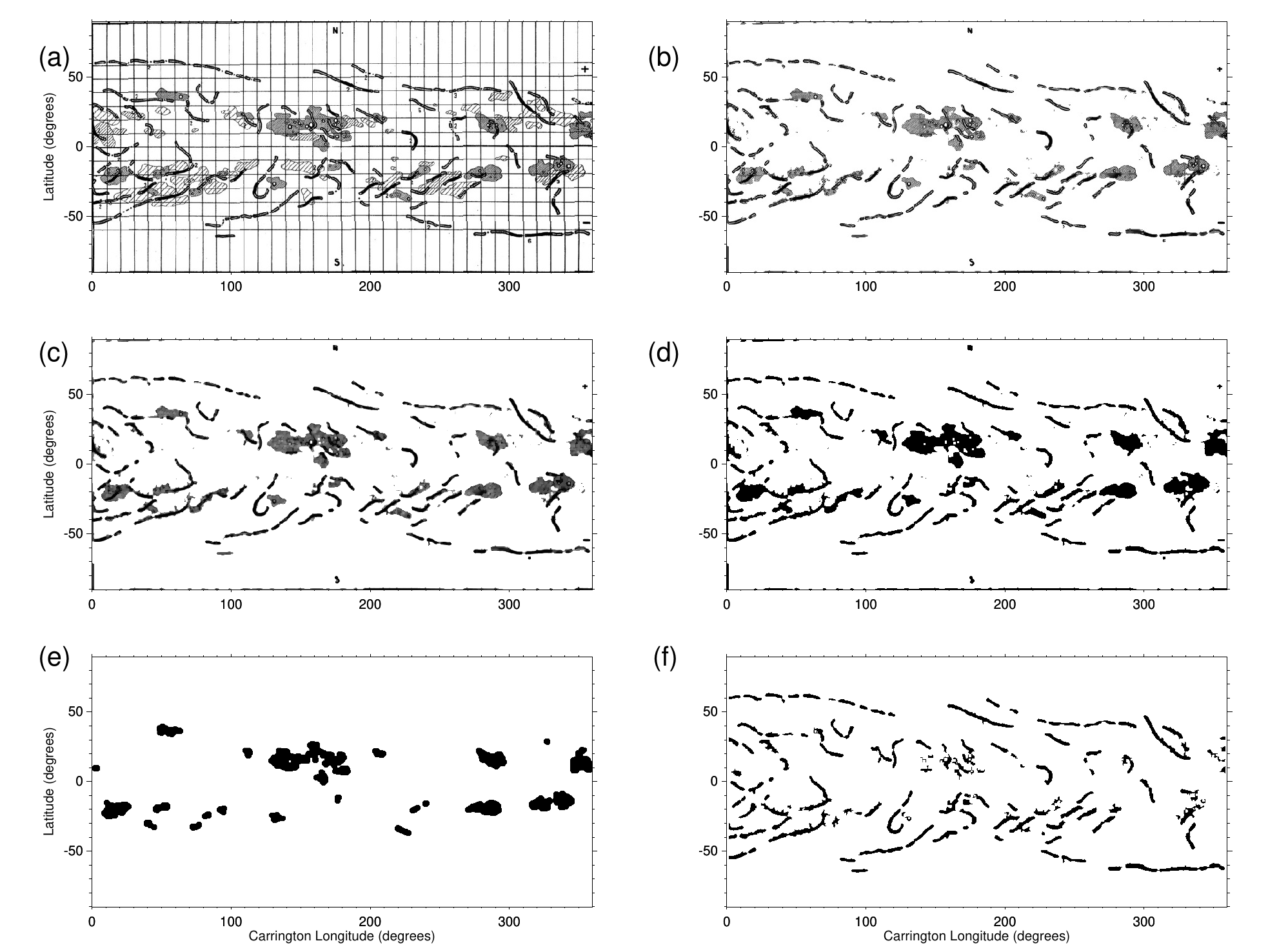}
\caption{Steps of filament detection Meudon hand-drawn Synoptic maps before rotation 1823. (a) Calibrated rotation 1678; (b) Map after removing grids; (c) Map after gray scale morphological closing of shaded structures; (d) Segmented image consisting of filaments and plages; (e) Image consisting of only plage; (f) Final image of detected filaments after removing plages. }
\label{detect}
\end{figure}

\begin{enumerate}
    \item We perform smoothing on the grid regions (detected from calibration steps) with 5$\times$5 median filter and replace all pixels lying above an intensity of 160 in the calibrated maps (Figure~\ref{detect}a) with value 255. This step retains regions with low enough intensity including filaments (Figure~\ref{detect}b).
    \item We perform grayscale morphological opening on the Step-1 image to join the hatched patterns inside filaments and create continuous structures (Figure~\ref{detect}c).
    \item We apply an intensity threshold of (median$-1.5\sigma$) producing a segmented image of filaments and plages (Figure~\ref{detect}d). $\sigma$ stands for the standard deviation of the entire Step-2 image (Figure~\ref{detect}c).
    \item We perform morphological opening to remove all the elongated regions (i.e. filaments) with kernel width of typical filament thickness. This step results in an image containing plages (Figure~\ref{detect}e).
    \item Subtracting Step 4 image from Step 3 image, subsequently retaining connected regions of an area greater than 20 pixels and regions within plage regions with intensity less than (median$-\sigma$) produce the final filament segmented image (Figure~\ref{detect}f). Here median and standard deviation ($\sigma$) are calculated over individual plage regions in Step 2 image (Figure~\ref{detect}c).
\end{enumerate}
Through the above detection method we detect a total 66,253 filaments from  948 Carrington maps (rotation 876-1823) available as black and white images in the Meudon database during the period 1919-1989.

\vspace{-0.005\textwidth}
\subsection{Rotation 1824-1853}
To detect the filaments in this rotation regime, we extract the pixels having intensity 4 (Figure~\ref{detect2}a). Subsequently, we fill those filaments found through morphological closing with a $5\times5$ kernel to produce the final filament segmented image (Figure~\ref{detect2}b).
\vspace{-0.01\textwidth}
\subsection{Rotation 1931-1992}
Filament information, for this range of rotation, are present in three different pixel values which are 1, 2 and 3. Equating the calibrated image (Figure~\ref{detect2}c) to values of 2, 3 we find most of the filament pixels. We extract rest of the filament pixels by equating the calibrated image to a value of 1 and filtering out areas less than or equal to 5 pixels. This step is necessary as filtering with just intensity value also results in grid locations. We subsequently fill those filament structures through morphological closing with a $5\times5$ kernel to produce final filament segmented image (Figure~\ref{detect2}d). 
\vspace{-0.005\textwidth}
\subsection{Rotation 1993-2008}
\begin{figure}
\centering
\includegraphics[scale=0.93,angle=0]{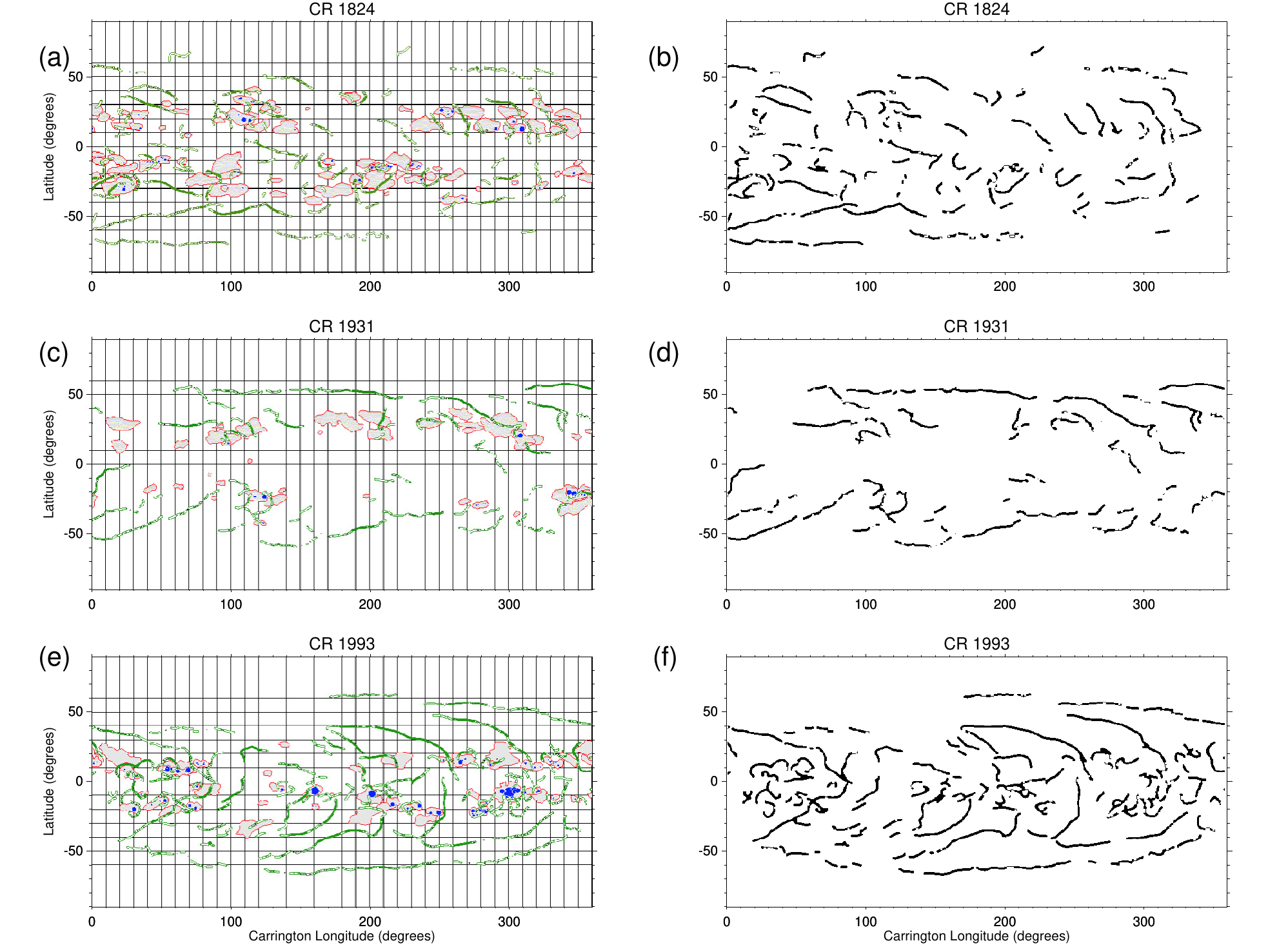}
\caption{Detection of filaments from Meudon maps for rotation 1824 onward. a) Calibrated maps for rotation 1824; b) Detected filaments from (a); c) Calibrated maps for rotation 1931; d) Detected filaments from (c); e) Calibrated maps for rotation 1993; f) Detected filaments from (e).}
\label{detect2}
\end{figure}

Extraction of filaments in this rotation regime is quite staright forward.We extract the green color-plane from the calibrated image (Figure~\ref{detect2}e).From the green color-plane, we extract pixel locations having intensity value of 153. We then fill those regionsthrough morphological closing with a $5\times5$ kernel to produce the final filament segmented image (Figure~\ref{detect2}f).

We extract a total of 22,065 filaments from 105 Carrington maps (rotation 1824-2008) available as colour images in Meudon database during the time period 1989--2003. Thus we extract 83,318 filaments in total from 1053 Carrington maps (rotation 867-2008).



All Carrington maps in McIntosh archive, being digitally color-coded, provides a straight forward way to extract filaments as presented in \citet{2018ApJ...868...52M}.

\begin{figure}
\centering
\includegraphics[scale=0.9,angle=0]{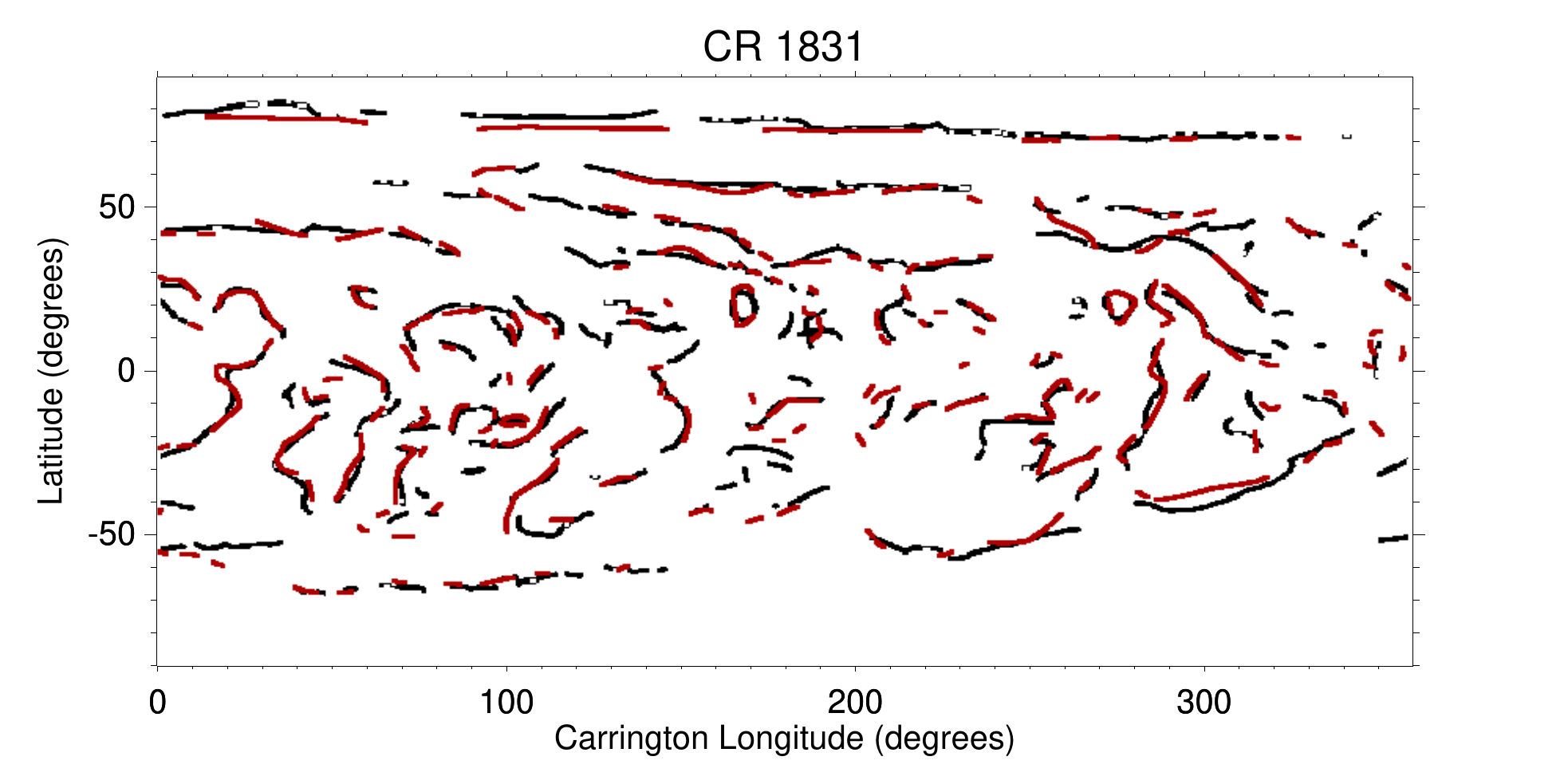}
\caption{Filament detected from Meudon CR 1831 (black) overplotted on those from McIntosh (red) for the same roration.}
\label{match}
\end{figure}

In Figure~\ref{match} we plot a Carrington map with detected filament from Meudon CR 1831 and ovelay the filaments of same rotation from McIntosh database. One can observe the visual match of individual filaments indicating consistent information from the two presented archives and also validating our methods of detection. Using the detected filaments, we extract different filament parameters, perform the statistical analysis and study corroboration of the two hand-drawn datasets as depicted in the following sections.

\pagebreak

 \section{Estimating Length and Tilt of the filaments}

We calculate filament length using the following formula \cite{2018ApJ...868...52M}.

\begin{equation}
\mathbf{ L = \sum_{n} \sqrt{R_{\odot}^{2}\delta\theta^{2}+R_{\odot}^{2}Cos^2 \theta \delta\phi^{2}}}
\end{equation}

\noindent where L is filament's length, $R_{\odot}$ is the solar radius, the symbols $\theta$ and $\phi$ represent the latitude and longitude of a particular pixel respectively and $n$ being the total number of pixels associated with the filament's structure. The quantities $\delta \theta$ and $\delta \phi$ are latitudinal and longitudinal differences between two adjacent pixels, respectively.
Then we add all filaments length in a Carrington map to get the total length of all filaments in that Carrington map.

We fit straight lines to each filament to estimate the filament tilt angle. For a particular filament, the angle of the inclination of the fitted straight line makes to the equator gives the measure of the tilt angle of that filament. We take the angle is positive in anticlockwise direction from the equator and negative in the clockwise direction from the equator.

\begin{figure}
\centering
\includegraphics[scale=1.1,angle=90]{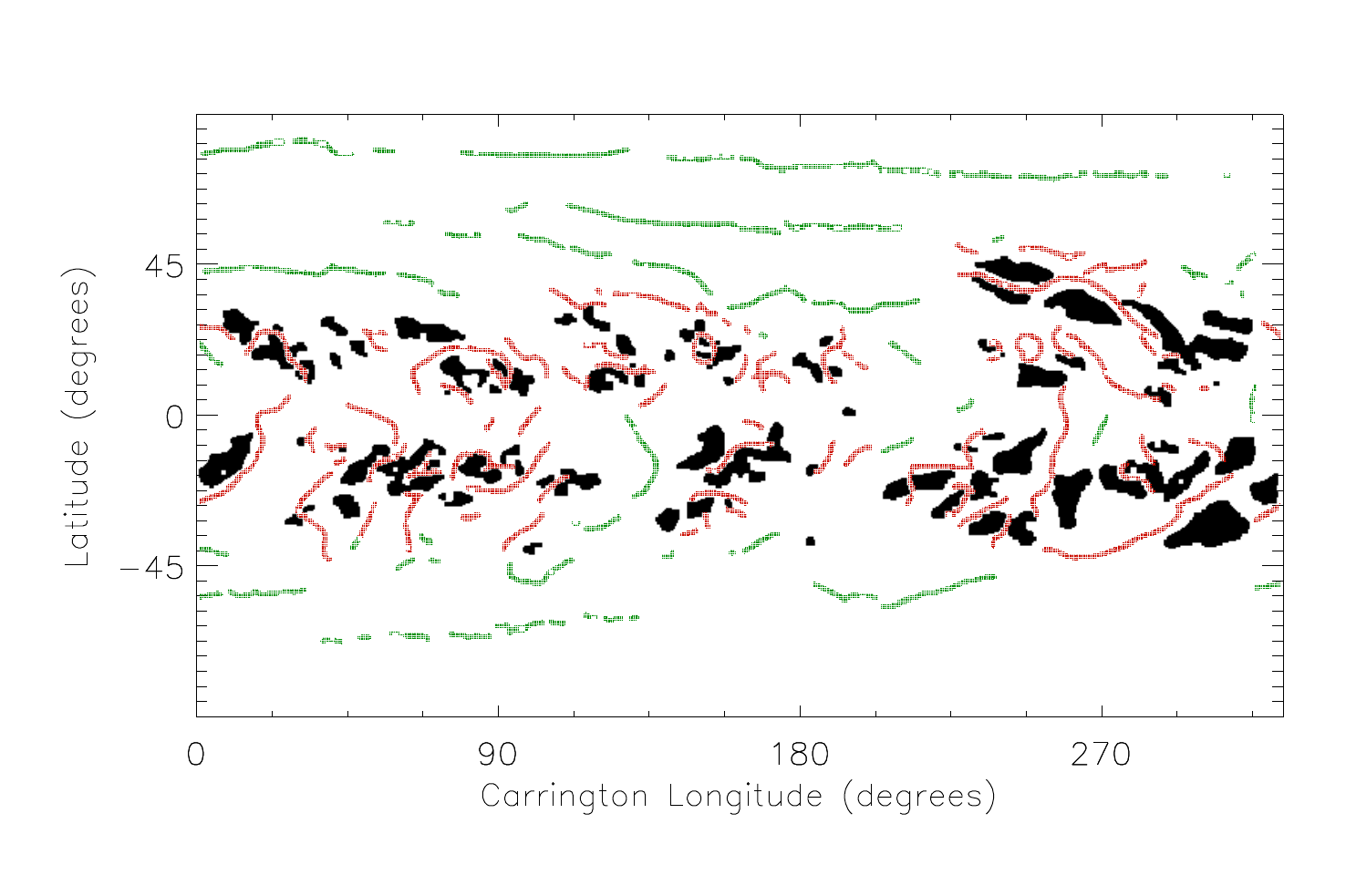}
\caption{Active region associated (marked in red) and unassociated (marked in green) filaments detected from Meudon CR 1831 (black) based on their distances from plages (marked in black).}
\label{association}
\end{figure}
\section{Identification of Active region association of filaments}
To ascertain the association of the filament with an active region we use the plages as identified  in the meudon maps. We define active region associated filaments to be those which are within 5 degree distance from any plage. Likewise we define active region unassociated filaments to be those which has greater than 5 degree distance from all the plages present in the map. To demonstrate this we show Figure~\ref{association}, Carrington map 1831 with detected plages (represented by black patches) and filaments (with red or green lines). By red lines we represent the active region associated with filaments, whereas the green lines  represent the active region unassociated filaments. This method of defining active region associated and unassociated filaments is an improvement to that of \cite{2018ApJ...868...52M}, in which association was defined only by latitude criterion. They defined active region associated filaments to be the filaments which are within the active region belt defined only by latitude extension (see the butterfly diagram of Figure 4 of \cite{2018ApJ...868...52M}) and thus overestimates the active region associated filaments which are within the active region belt but separated by longitude. Thus in addition to a longer time length of the Meudon data-set our improved method of estimating active region association of filaments  gives better confidence to the statistical results which crucially depends on the number of active region associated and unassociated filaments
\pagebreak

\section{Results and Discussion}\label{res_disc}

\subsection{Total Filament Number and Total Filament Length Variation}
Solar activity varies periodically with a period of 11 years. The filament number also varies periodically \citep{2015ApJS..221...33H,2016SoPh..291.1115T,2019RAA....19...80M}. The total filament length gleaned from Carrington maps are also known to vary with 11 year periodicity \citep{2016SoPh..291.1115T,2018ApJ...868...52M}.

\begin{figure}
\centering
\includegraphics[scale=1.25,angle=90]{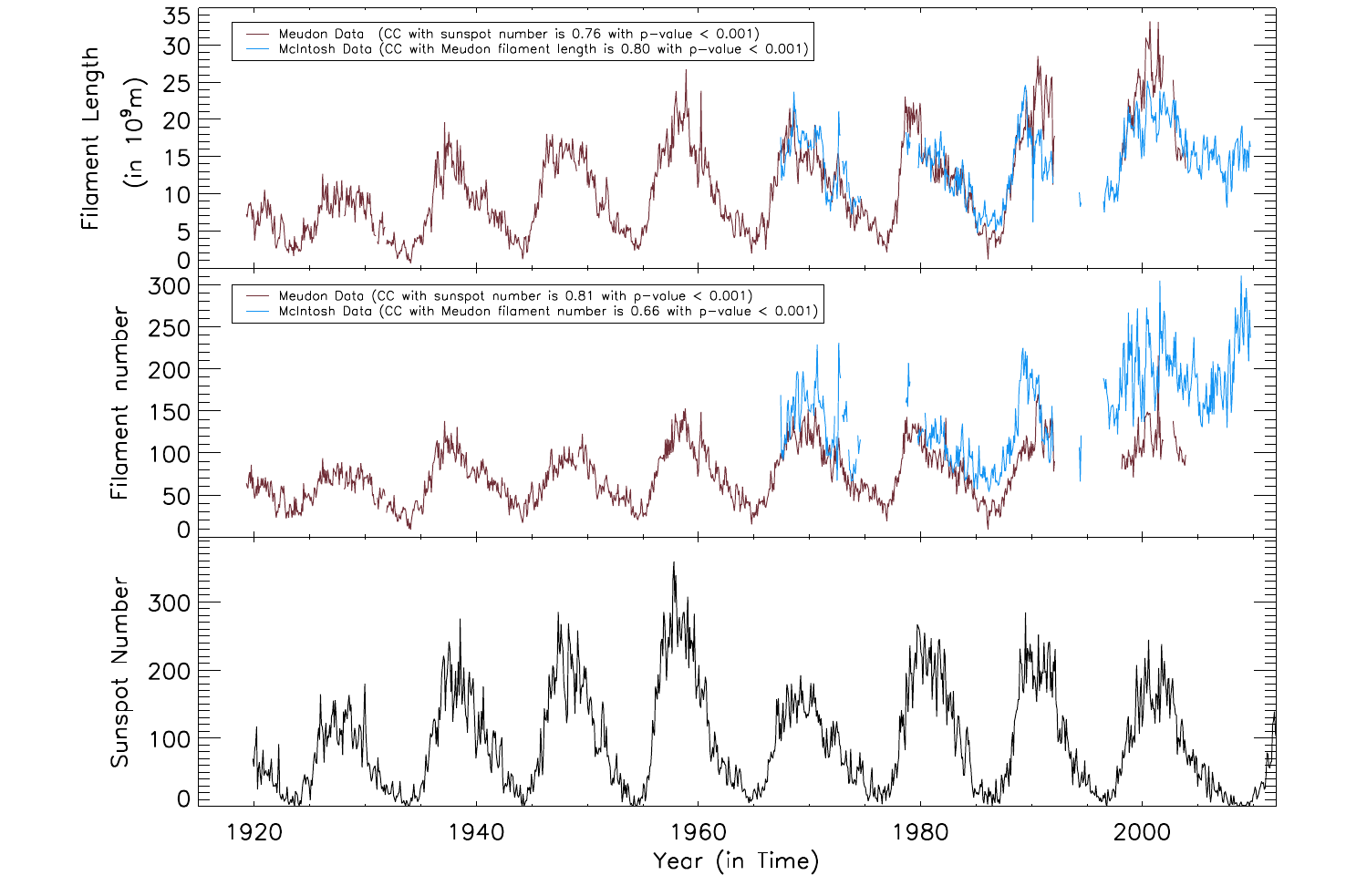}
\caption{Lon-term variation of filament number, filament length, and sunspot number. The top panel shows the variation of total filament Length ($L_{tot}$) with time. The brown line represents the variation of total filament length with time from Meudon data and the blue line represents the variation of total filament length with time from McIntosh data. The middle panel shows the variation of total filament number ($N_{tot}$) with time. The brown line represents the temporal variation of total filament number  from Meudon data and the blue line represents the same from McIntosh data. In the bottom panel, the black line represents the sunspot number cycle with the data taken from Solar Influences Data Center (SIDC, \url{http://www.sidc.be/silso/home}).}
\label{fil_length_var}
\end{figure}

We count all the filaments in each Carrington map and define it as a total filament number corresponding to that Carrington map ($N_{tot}$) to study the variation of filament numbers with time. The middle panel of  the Figure~\ref{fil_length_var} depicts the variation of total filament number ($N_{tot}$) with time. The brown line represents the variation of total filament number with time from Meudon data and the blue line represents the variation of total filament number with time from McIntosh data. We find the for the overlapping period (1967 July- 2003 October) the variation of the total filament of Meudon data is correlated with variation of total filament number of McIntosh data with Spearman rank correlation coefficient of 0.66 and confidence of $99.99\%$. We add all filaments' length in a Carrington map in a Carrington map and define it as total filament length ($L_{tot}$) for that Carrington map to study the variation of total filament length with time. The top panel of  the Figure~\ref{fil_length_var} depicts variation of $L_{tot}$  with time. The brown curve represents the variation of total filament length from Meudon data and the blue curve represents the variation of total filament length from McIntosh data. There is an overlap of data in cycle 20-23. We found a good match in the variation of filament length between the Meudon and McIntosh databases in the overlapping period. We find that for the overlapping period the variation of total filament length from Meudon data is correlated with variation of total filament length of McIntosh data with Spearman rank correlation coefficient of 0.80 with the confidence of $99.99\%$. In the bottom panel, the black curve represents the variation of sunspot numbers with time. The periodic variation of both filament number and total filament length is in phase with the sunspot cycle. We find the sunspot number variation is correlated with  Meudon total filament numbers ($N_{tot}$) variation with a  Spearman rank correlation coefficient of 0.81 with the confidence of $99.99\%$. The sunspot number variation is correlated with  Meudon total filament length ($L_{tot}$) variation with a  Spearman rank correlation coefficient of 0.76 with a confidence of $99.99\%$ (see Figure~\ref{fil_length_var}).

\subsection{Latitude and Longitude distribution of filament}
\begin{figure}
\centering
\includegraphics[scale=1.0,angle=90]{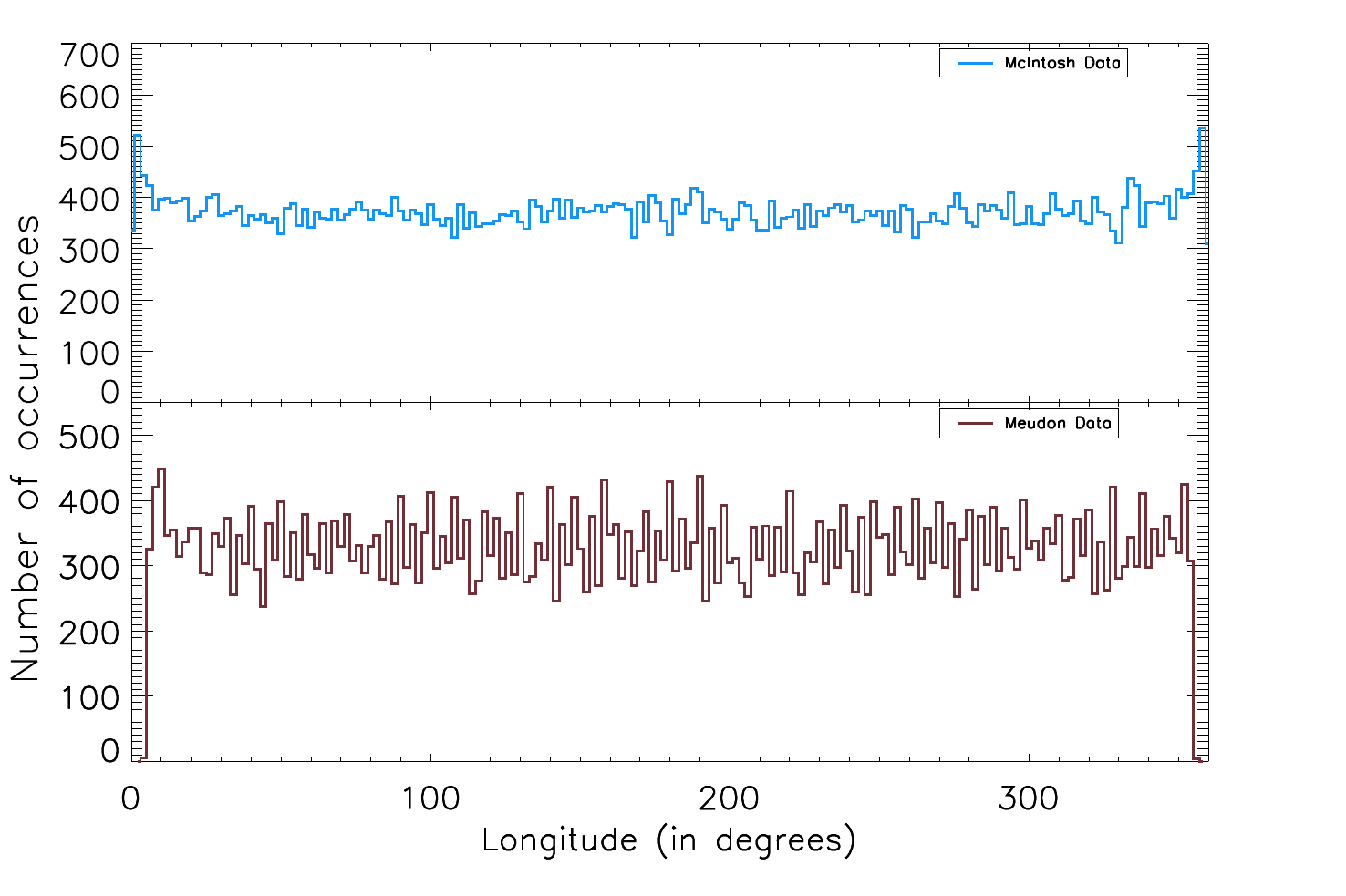}
\caption{Histogram of longitude distribution of filament centers. The upper  panel shows the histogram of filament longitude from McIntosh data (in blue). The lower  panel shows the same for Meudon database (in brown).}
\label{long_hist}
\end{figure}

\begin{figure}
\centering
\includegraphics[scale=1.3,angle=90]{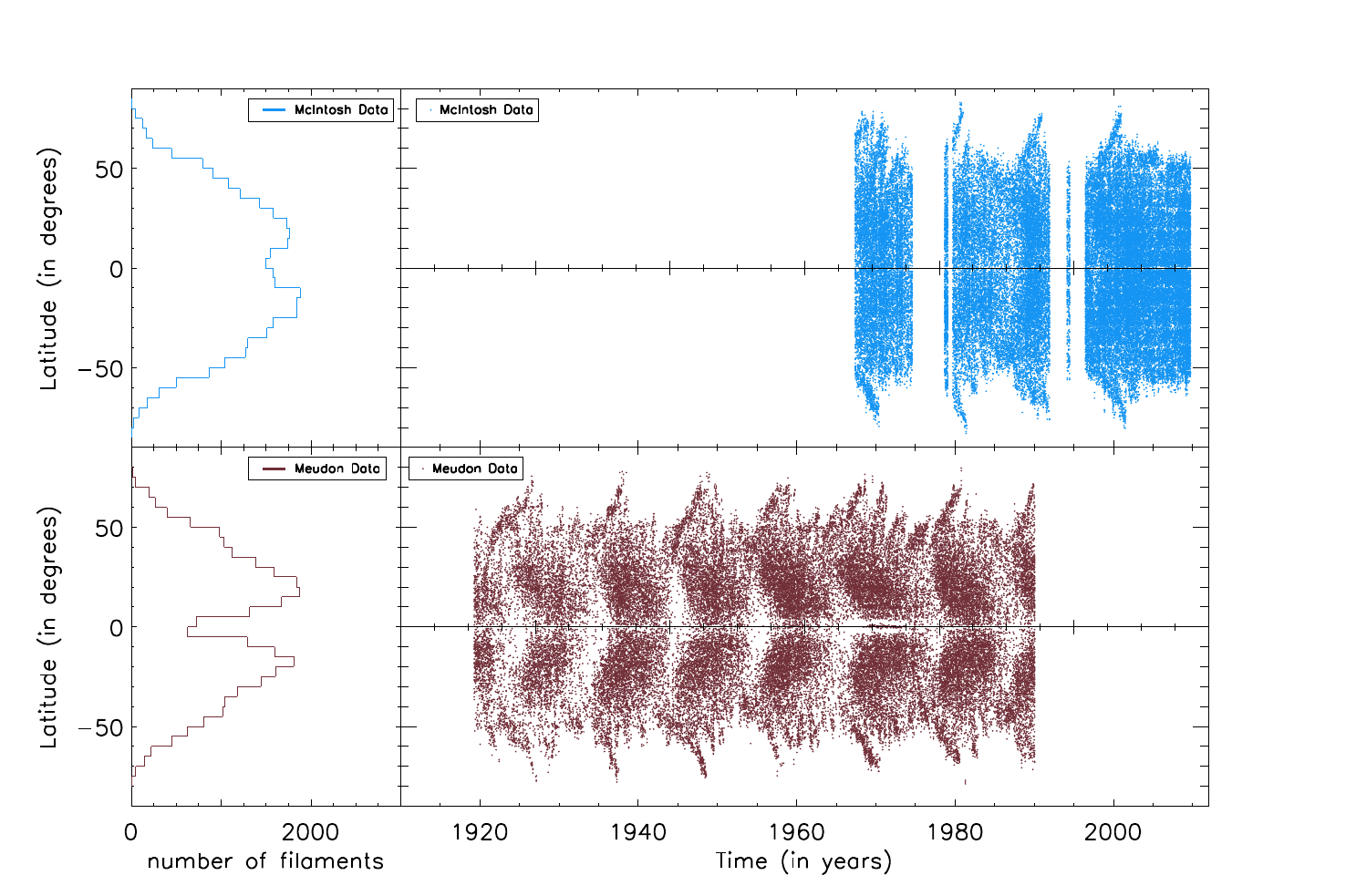}
\caption{Temporal variation of latitudinal distribution of the filaments centers. The upper left panel shows the histogram of filament latitude from McIntosh data (in blue). The lower left panel shows the same for Meudon data (in brown). The upper right panel depicts the temporal variation of filaments' latitude (the blue dots) from McIntosh data. The lower right panel shows the temporal variation of filaments' latitude (the brown dots) from Meudon data.}
\label{lat_temp_dits_md}
\end{figure}

The distribution of filaments in longitude is uniform \citep{2018ApJ...868...52M} in the solar disk (see Figure~\ref{long_hist}). However, the distribution of filaments in latitude is not uniform. The left panels in Figure~\ref{lat_temp_dits_md} depicts the histogram of filaments in latitude from the McIntosh database (left top panel) and the Meudon database (left bottom panel). We find a bimodal distribution with a peak between $10^{o}$ to $30^{o}$, which is similar to the observation by \cite{2018ApJ...868...52M} from the McIntosh database and \cite{2015ApJS..221...33H} from the GONG database. 

The right panels in Figure~\ref{lat_temp_dits_md} show the temporal variation of latitudinal distribution of the centers of filaments. We see a butterfly structure similar to the sunspot butterfly diagram. However, the latitudinal extent of the sunspot butterfly diagram is spread within $\pm 40^{o}$ whereas in case of filaments we see a larger latitudinal spread. We observe a poleward migration during solar cycle maxima. This poleward migration is popularly known as the ``rush to the pole" phenomena. These results are in agreement with earlier studies \citep{2016SoPh..291.1115T,2018ApJ...868...52M}. \cite{2015ApJS..221...33H} have reported similar butterfly patterns in the temporal variation of the latitudinal distribution of filament centers from the Big Bear Solar Observatory (BBSO) archive. \cite{2017ApJ...849...44C} have also reported a similar pattern analyzing the Kodaikanal archive. \cite{2018ApJ...868...52M} observed that the poleward migration of filaments coincides with the polar field reversal. The timing of polar field reversal is crucial information necessary for predicting the next solar cycle. However, observations of the polar field are challenging due to debilitating projection effects. So the timing of poleward migration of filaments can be used as a proxy for polar field reversal setting a crucial constraint to solar cycle evolution. 

\begin{figure}
\centering
\includegraphics[scale=0.6,angle=0]{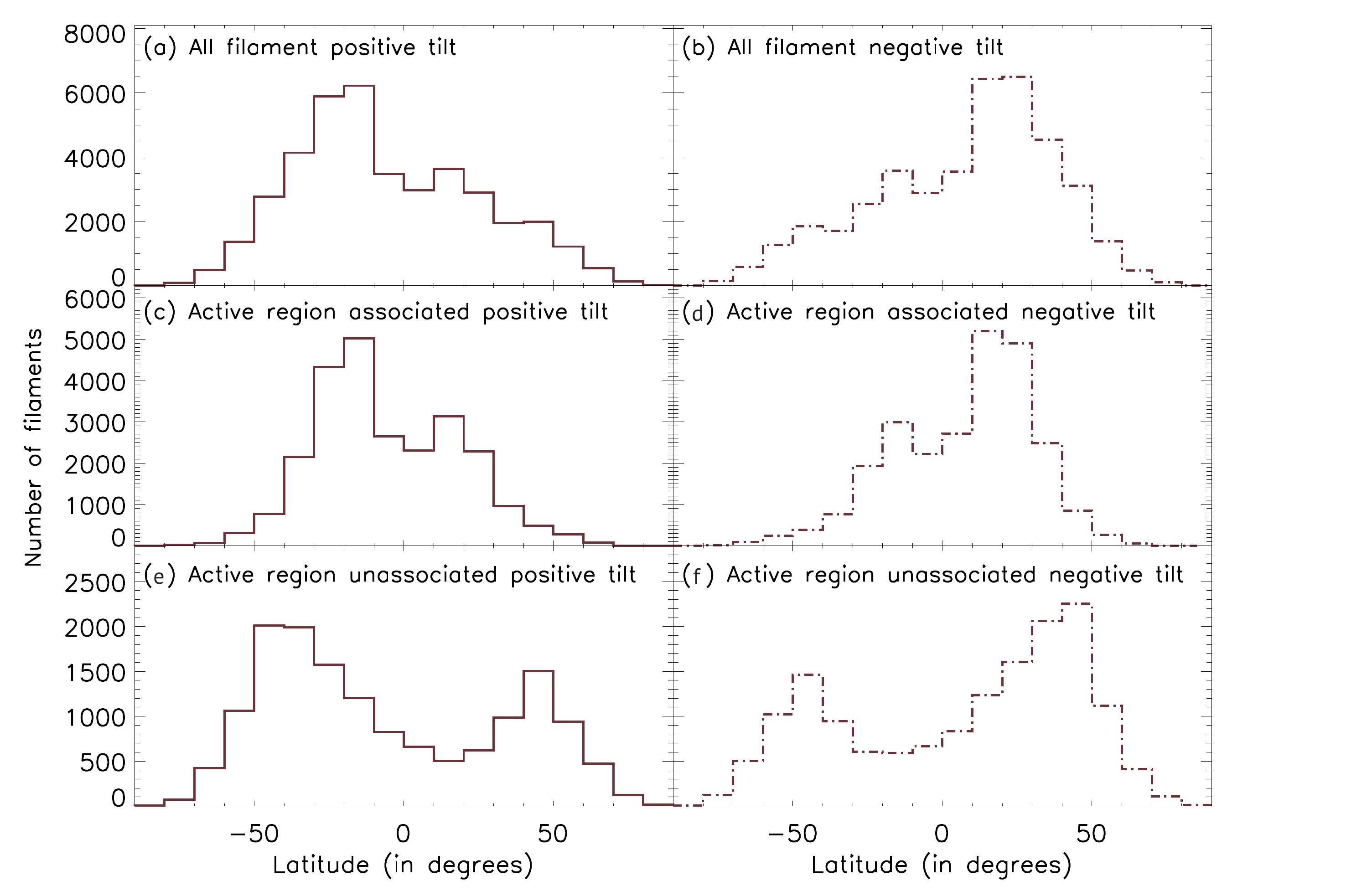}
\caption{Histogram of filament's latitude of filaments having either only positive tilt or only negative tilt.
(a) histogram of the latitude of all filaments with positive tilt; (b) represents the histogram of the latitude of all filaments with a negative tilt; (c) depicts the histogram of the active region associated filament with positive tilt; (d) depicts the histogram of the active region associated filament with a negative tilt; (e) depicts the histogram of active region unassociated filament with positive tilt; (c) shows the histogram of active region unassociated filament with a negative tilt.}
\label{fil_hist_all}
\end{figure}
 \begin{figure}[!htbp]
\centering
\includegraphics[scale=0.5,angle=90]{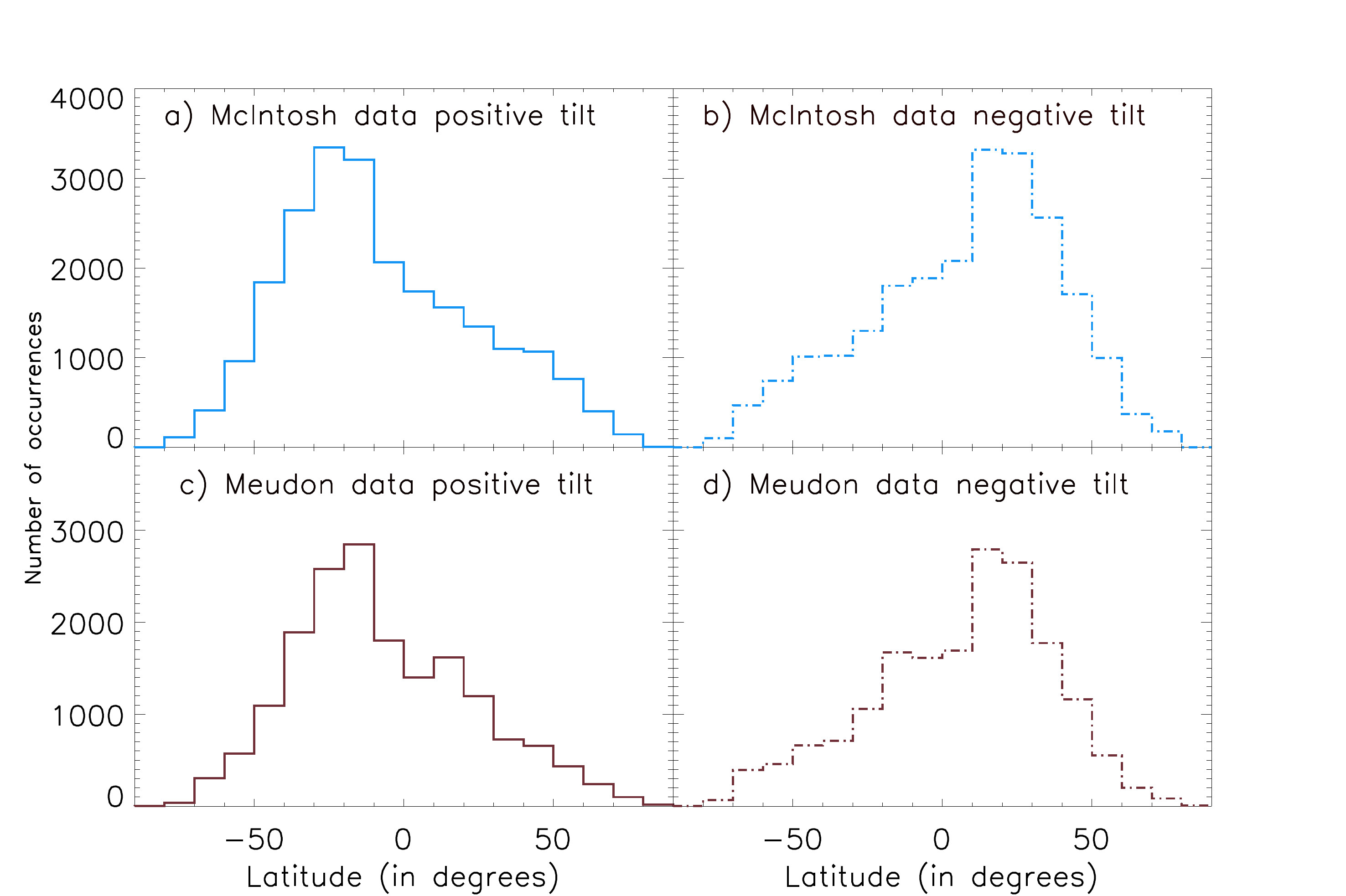}
\caption{Histogram of filament's latitude of filaments having either only positive tilt or only negative tilt in the common era (1967 April - 2003 October) of McIntosh and Meudon database. (a) and (b) represent the histogram of the latitude of filament from McIntosh data for the common era with positive tilt and negative tilt only respectively.
(c) and (d) represent the histogram of the latitude of filament from Meudon data for common era with positive tilt and negative tilt only respectively.}
\label{fig_tilt_common_era}
\end{figure}
\subsection{Tilt angle analysis of filaments}
We analyze the tilt angle of the filaments to find their origin. We segregate the filaments as an active region associated filaments and active region un-associated filaments depending on their distance from plages.

\begin{figure}[!htbp]
\centering
\includegraphics[scale=0.25,angle=0]{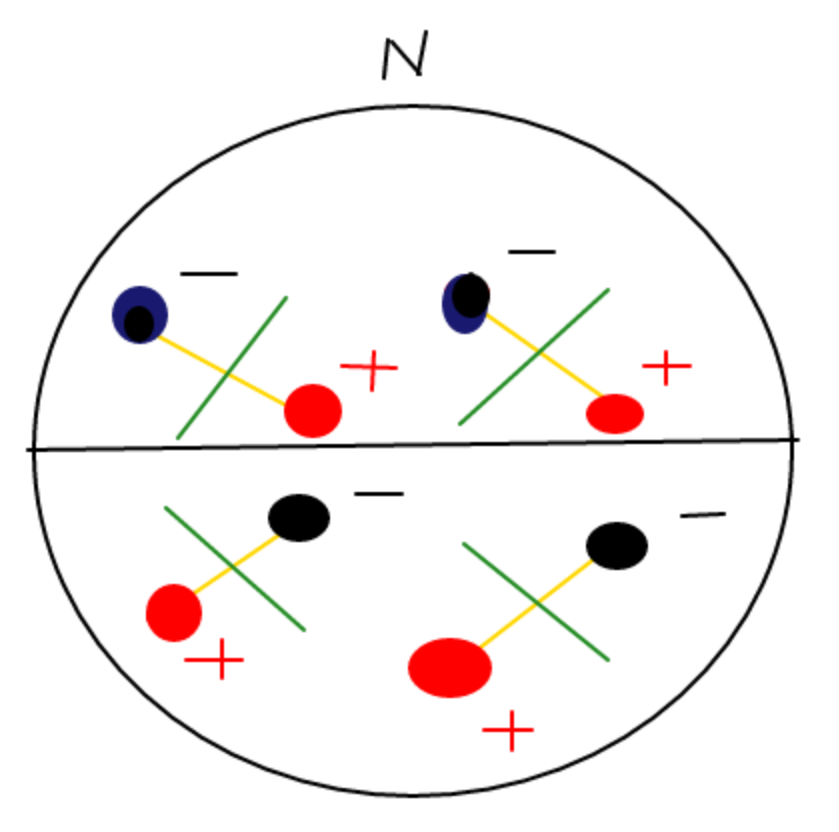}
\caption{Cartoon of a tilt angle of the active region (yellow lines) and associated polarity inversion line (green lines). The red filled circle with plus sign depicts the positive polarity of the active region and the black filled circles with negative sign depict the negative polarity of the active regions. The N denotes the Northern hemisphere of the Sun.}
\label{fil_cartoon}
\end{figure}

Figure~\ref{fil_hist_all}.(a) depicts the latitude distribution of all positive tilt angle filaments. The southern hemisphere filaments tend to have excess positive tilt in comparison to the northern hemisphere. Figure~\ref{fil_hist_all}.(b) depicts the latitude distribution of all negative tilt angle filaments. Northern hemisphere filaments have an excess negative tilt in comparison to the southern hemisphere. Figure~\ref{fil_cartoon} depicts a cartoon of an active region and its associated Polarity Inversion line. We follow the convention that the tilt angle of active regions in the northern (southern) hemisphere is negative (positive) according to Hale's polarity law \citep{1919ApJ....49..153H}. Now the polarity inversion lines and their tilts  are perpendicular to active region tilts. Hence tilt angles of Polarity inversion line associated filaments should be positive in the northern hemisphere and negative in the southern hemisphere. The yellow  lines joining the opposite polarities (positive polarities represented by red circles and negatives by black) in Figure~\ref{fil_cartoon} represents tilt of active regions which is evidently positive in the northern hemisphere and negative in the southern hemisphere. Whereas active region polarity inversion lines are depicted by green lines, which being perpendicular to the active region tilt (depicted by yellow lines) follows exact opposite behaviour to those of active region tilt. So polarity inversion line tilts (represented by green lines) are expected to be  positive in the northern hemisphere and negative in the southern hemisphere. Filament usually forms along the Polarity inversion lines \citep{1972RvGSP..10..837M,1983SoPh...85..215M,1998SoPh..182..107M}.Thus active region associated filaments should follow the same tilt angle pattern as active region polarity inversion lines. So we should expect a positive tilt for active region associated filaments in the northern hemisphere and a negative tilt  in the southern hemisphere. However the active region unassociated filaments  may or may not follow this pattern since they are known to be formed along polarity inversion lines of large scale magnetic field in solar disk and not constrained within active regions. Now Figure~\ref{fil_hist_all}.(a) shows southern hemispheric filaments to have excess positive tilt in comparison to the northern hemisphere  Figure~\ref{fil_hist_all}.(b) depicts northern hemisphere filaments have an excess negative tilt in comparison to the southern hemisphere; exactly opposite pattern to those expected from active-region associated filament. However since the complete filaments distribution includes active region unassociated filaments so we segregate the filaments as active region associated filaments and active region unassociated filaments (the methods of segregation is described in section 6) to see  whether contribution from active region unassociated filaments produces this anomaly in the overall filament tilt angle distribution.

Figure~\ref{fil_hist_all}.(c) represents the histogram of the latitude of all positive tilt angles of active region associated filaments. Even in this set, we notice that the southern hemisphere dominates a positive tilt angle. Figure~\ref{fil_hist_all}.(d) represents the histogram of latitudes of all negative tilt angle filaments located near the active region belt. We notice that the northern hemisphere has a dominance of negative tilt angle filaments even in this set. Figure~\ref{fil_hist_all}.(e) represents the histogram of the latitudes of all positive tilt filaments which are not associated with active regions. In this sample, the southern hemisphere dominates positive tilt angle filaments. Figure~\ref{fil_hist_all}.(f) represents the histogram of latitudes of all negative tilt angle filaments not associated with active regions. The northern hemisphere dominates negative tilt angle filaments.

Thus both the sub-set filaments, either located near active regions or located away from active regions, have a tilt angle which is opposite (similar anomaly has been reported by \cite{2018ApJ...868...52M} and \cite{2019RAA....19...80M} from McIntosh database) to what would be expected if they are formed on the PILs between the two spots of an individual active region.

\subsection{Towards a composite uniform long term data-set of filaments}
Figure~\ref{fig_tilt_common_era} depicts a histogram of latitude of positive and negative tilt filaments similar to Figure~\ref{match} but only for the co temporal data where both data sources, Meudon and McIntosh have records.  Together with Figure~\ref{match} which shows a one to one co-spatial relation of filaments, these similar tilt angle distribution from both the data-set provides robustness of our method of detection of filaments from hand-drawn Meudon data and  pave  a  pathway  towards  building  a  composite  data  of  solar  filaments.  This  will  work  as  a unique resource to solar physicists for the study of long term variation in solar filaments

 \begin{figure}[!htbp]
\centering
\includegraphics[scale=.8,angle=90]{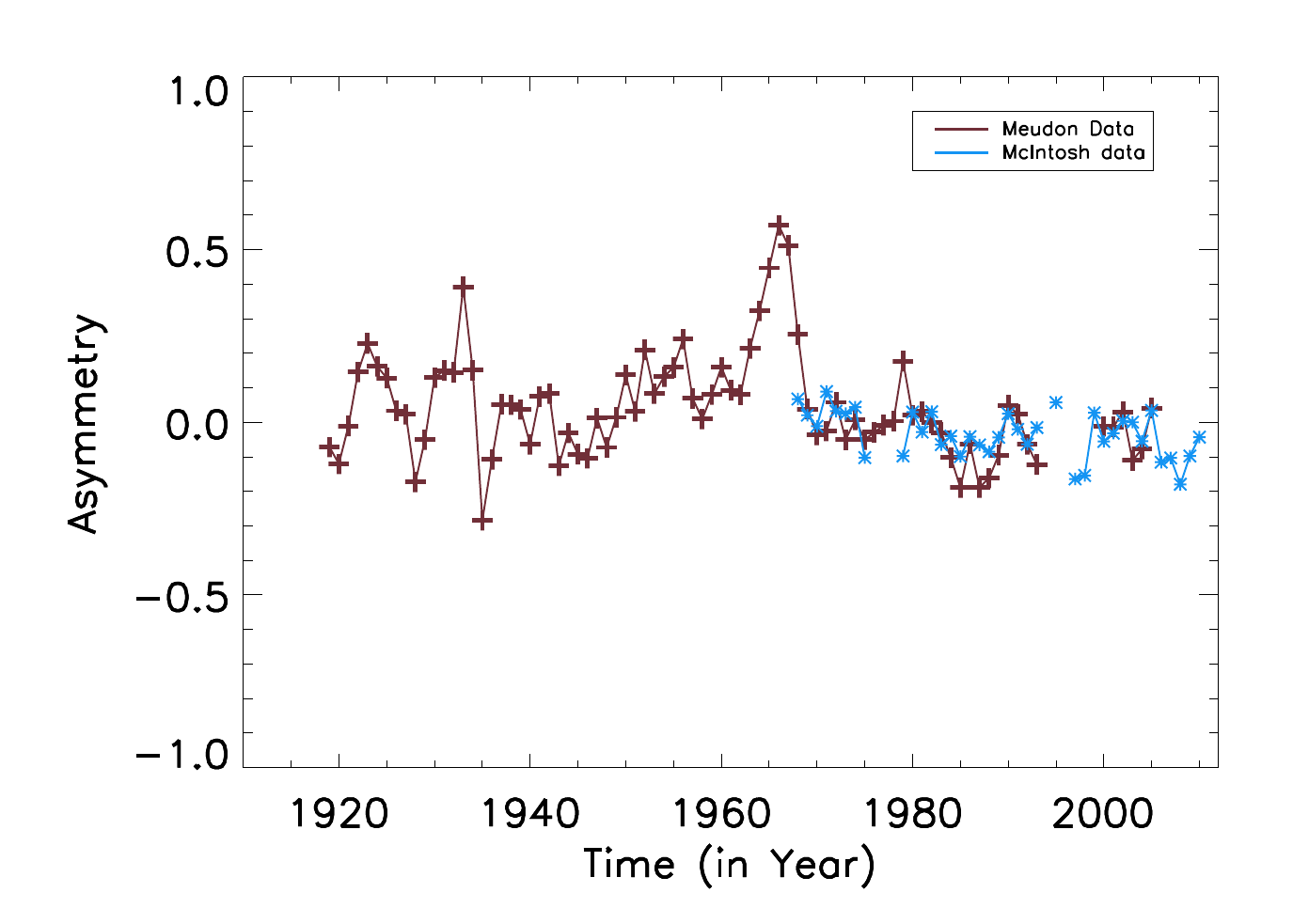}

\caption{Asymmetry of filament number. The brown line shows the temporal variation of asymmetry in filament number from Meudon database. The blue line depicts the same from McIntosh database.}

\label{fig_asymmetry}
\end{figure}

\section{Asymmetry}
The generation of magnetic field in the Sun is not symmetric in two hemispheres which introduces an asymmetry in observed solar magnetic features. 



Earlier studies \citep{2018ApJ...868...52M,2019RAA....19...80M,2015ApJS..221...33H} have found that the generation of filaments is not symmetric in the two hemispheres.
We define the asymmetry in filament numbers $A_{f}$ to be
\begin{equation}
A_{f}=\frac{N_{f}-S_{f}}{(N_{f}+S_{f})},
\end{equation}

\noindent where $N_{f}$ and $S_{f}$ are the total filament numbers in the northern and southern hemispheres, respectively. So, if $A_{f} > 0$, implies that the total filament number in the northern hemisphere is more in comparison to the southern hemisphere and $A_{f} < 0$, implies that total filament number in the southern hemisphere is more in comparison to the northern hemisphere. The brown line in 
Figure~\ref{fig_asymmetry} represents the variation of asymmetry in filament numbers from Meudon database. The blue line in 
Figure~\ref{fig_asymmetry} represents the variation of asymmetry in filament numbers from McIntosh database. The filament number asymmetry, $A_{f}$ fluctuates over the time of observation with a prominent inclination towards the northern hemisphere during the time 1919 to 1966 (which is covered by the Meudon data only). After 1966 again the $A_{f}$ fluctuates but this time with dominance on the southern hemisphere, which is in agreement with an earlier report by \cite{2018ApJ...868...52M}.  We find that the north-south asymmetry of filament numbers from Meudon data is correlated with the north-south asymmetry of filament numbers from McIntosh data during overlapping periods with a Spearman rank correlation coefficient 0.50 with confidence of $99.99\%$.

\section{Summary and Conclusions}\label{concl}

We develop an automated method 
to calibrate and detect filaments from Meudon hand-drawn Carrington maps both in black/white and coloured formats. We extract 83,318 filaments in the Meudon database during the period 1919 to 2003 using our detection method. We compare our prior results of filament detection from the McIntosh database \citep{2018ApJ...868...52M} during the overlapping period (1967 April to 2003 October) and find a good agreement. We determine the tilt angle of the filaments to explore their origin. We perform a detailed analysis of the properties of filaments and their evolution with time over solar cycles 15-23. 
 
Solar activity varies periodically. Similar to sunspot number and sunspot area, the filament number \citep{2010NewA...15..346L,2015ApJS..221...33H,2016SoPh..291.1115T,2019RAA....19...80M} and total filament length \citep{2015ApJS..221...33H,2016SoPh..291.1115T,2018ApJ...868...52M} also varies periodically. From the Meudon observations, currently the longest available baseline of hand-drawn filaments, we confirm that filament number and the total filament length vary periodically in phase with the sunspot number variation. We find the sunspot number variation is correlated with  Meudon total filament numbers variation with a Spearman rank correlation coefficient 0.81 with the confidence of $99.99\%$ .
We find the variation of filament number and filament length in the overlapping period (solar cycle 20-23) with the McIntosh database to be consistent. We find in the overlapping period the variation of total filament length from Meudon data is correlated with variation of the same from McIntosh data producing a Spearman rank correlation coefficient of 0.80 with the confidence of $99.99\%$.  We also find the sunspot number variation to be correlated with the Meudon total filament length variation with a  Spearman rank correlation coefficient 0.76 with the confidence of $99.99\%$. 

The filament distribution in longitude is uniform \citep{2018ApJ...868...52M} but the latitudinal distribution of filaments shows a bimodal nature \citep{2015ApJS..221...33H,2018ApJ...868...52M}. We reconfirm from the Meudon database that the longitudinal distribution of filaments is indeed uniform and the latitudinal distribution shows a bimodal distribution with peaks between 20$^{\circ}$ and 30$^{\circ}$. The temporal variation of the latitudinal distribution of filaments shows a butterfly structure similar to earlier reports \citep{1948AnnObsParis...6..7,1994A&A...290..279M,2015ApJS..221...33H,2016SoPh..291.1115T,2018ApJ...868...52M}. The spread of the latitude in case of filaments is larger in comparison to the latitudinal spread of sunspots in the butterfly diagram. Similar to the results of the earlier reports \citep{2015ApJS..221...33H,2016SoPh..291.1115T,2018ApJ...868...52M} we also observe a migration of the filaments to the pole during cycle maxima in the Meudon database.

According to Hale's polarity law \citep{1919ApJ....49..153H} tilt angle of the bipolar active region is negative in the northern hemisphere (with respect to the equator) and positive in the southern hemisphere. Since the tilt angle of polarity inversion line (PIL) is perpendicular to the active region tilt, the tilt of active region PIL carrying filaments is expected to be positive in the northern hemisphere and negative in the southern hemisphere. However, our study of Meudon data reveals the tilt angle of filaments -- both near and away from active regions -- to follow an opposite trend. The tilt is predominantly negative in the northern hemisphere and positive in the south. This indicates that filaments predominantly originate not within active regions PILs, but outside of it in the large-scale inter active region fields or even larger-scale surface magnetic field distributions. Our finding confirms the result by \cite{2018ApJ...868...52M} from McIntosh database. Thus our work puts the finding of the anomalous behaviour of filaments by \cite{2018ApJ...868...52M} from the McIntosh database in firmer ground suggesting filaments originate in the large-scale field distributions beyond individual active regions. This result is now confirmed from the longest filament database and need to be addressed for the understanding of the evolution of the large scale field on the sun.  

Furthermore, like other magnetic features in the Sun, filaments are also not symmetric in the two hemispheres and this asymmetry also varies with time \citep{2018ApJ...868...52M,2015ApJS..221...33H}. 
We find the north-south asymmetry of filament numbers from the Meudon database is correlated with the north-south asymmetry of filament numbers from McIntosh data during the overlapping period with a Spearman rank correlation coefficient 0.50 with the confidence of $99.99\%$. This asymmetry in filament distribution may arise from the underlying asymmetry of magnetic field emergence associated with the solar dynamo mechanism \citep{Passos2014,Bhowmik2019,Hazra2019} and (or) they might be related to hemispheric asymmetries in large-scale plasma flows \citep{Lekshmi2018, Lekshmi2019}; these merit further theoretical investigations.

Solar filament and prominence structures often get destabilized and cause eruptive events such as CMEs -- which can generate severe space weather \citep{Sinha2019}. However, direct observations of CMEs have existed only from the Solar and Heliospheric Observatory (SOHO)-LASCO (Large Angle Spectroscopic Coronagraph) era since 1996. In this context,century-scale filament time series such as ours may be utilised to reconstruct plausible eruptive space weather events in earlier times, well before the space age.
 
In summary, our work establishes the consistency of the properties and long term evolution of filaments gleaned from two major, hand-drawn filament archives. We would like to extend our analysis to other long term archives such as the Kodaikanal Solar Observatory data in the future. These archives may act as foundations of independent long-term filament studies, constrain models of solar magnetic field evolution, and act as an indicator of eruptive space weather events before direct CME observations associated with filament eruptions were possible.

\acknowledgements The Center of Excellence in Space Sciences India (CESSI) is funded by the Ministry of Education, Government of India under the Frontier Areas of Science and Technology (FAST) scheme. DB acknowledges support from the Department of Science and Technology (DST), Government of India through the Cluster Project titled "Preservation and analysis of the century long solar data from Kodaikanal Solar Observatory". 

\bibliography{references}
\bibliographystyle{aasjournal}

\end{document}